\documentclass{elsart}
\usepackage{amssymb}
\usepackage{amsmath}
\usepackage{amsfonts}

\font\frak=eufm10  scaled \magstep 1   
 1  

\def\QED{\hskip0.1em\hfill\null\ \null\nobreak\hfill
\kern3pt\lower1.8pt\vbox{\hrule\hbox   {\vrule\kern1pt\vbox{\kern1.7pt
\hbox{$\scriptstyle   QED$}\kern0.2pt}\kern1pt\vrule}\hrule}}

\def\dfrac#1#2{{\displaystyle{\frac#1#2}}}

\def\goth #1{\hbox{{\frak #1}}}
\def\pd#1#2{\frac{\partial#1}{\partial#2}}

\def\matriz#1#2{\left( \begin{array}{#1} #2 \end{array}\right)}

\def\<#1>{\langle #1\rangle}

\begin{document}

\begin{frontmatter}

\title{Nonlinear superpositions and Ermakov systems}

\author{Jos{\'e} F. Cari{\~n}ena, Javier de Lucas and Manuel F. Ra\~ nada}

\address{Departamento de  F\'{\i}sica Te{\'o}rica\\
Universidad de Zaragoza, 50.009 Zaragoza, Spain.}

\begin{abstract}
 The theory of superposition rules for solutions of a Lie system of first-order differential equations is extended to deal with analogous systems of second-order and 
the theory is illustrated with the very rich example of Ermakov-like systems.
\end{abstract}

\end{frontmatter}

\section{Introduction}

The characterization of non-autonomous systems of first-order differential
equations
\begin{equation}
\frac{dx^i}{dt }= Y^i(t,x)\,,\qquad  i=1,\ldots,n\,,\label{nonasys}
\end{equation}
admitting
a superposition rule is due to Lie \cite{LS}
and such a problem has being receiving very much attention during the last thirty
years because of its very important applications in physics \cite{NI1}-\cite{CarMarNas}.
The theory has recently been revisited from a more geometric approach in
\cite{CGM06}
where the r\^ole of the superposition function is played by an appropriate
connection.
The main point is that this new approach allows us to consider  partial
superposition of solutions as well and,  furthermore, it also allows  superposition of
 solutions of a
given system in order to obtain solutions of a new system. Our aim here is to
show how such a superposition rule may be understood from a geometric viewpoint
in a very simple but interesting case, the so called Ermakov-Pinney system
\cite{{ermakov},{P50}} as well as for other generalizations of it.  
 Such a system is made of second-order differential equations but
the theory developed by Lie can easily be adapted to deal with such SODE systems.
We find in this way room for implicit nonlinear superposition rules in the 
terminology of \cite{RR80b,{WS81}} and the so called Ermakov-Lewis invariants \cite{Le67}
appear in a natural way as functions defining the foliation associated to the
superposition rule. Moreover all reduction  techniques developed for Lie
systems \cite{CarRamGra,CarRamcinc} are also valid in these cases.

\section{Systems of differential equations admitting a superposition rule}

The superposition rule for solutions of (\ref{nonasys}) is  determined by a function 
$\Phi:{\mathbb{R}}^{n(m+1)}\to {\mathbb{R}}^n$,
\begin{equation}
x=\Phi(x_{(1)}, \ldots,x_{(m)};k_1,\ldots,k_n)\ ,\label{superpf}
\end{equation}
such that the general solution can be written, for
sufficiently small $t$, as
\begin{equation}
x(t)=\Phi(x_{(1)}(t), \ldots,x_{(m)}(t);k_1,\ldots,k_n)\ ,\label{superpft}
\end{equation}
where $\{x_{(a)}(t)\mid a=1,\ldots,m\}$ is a fundamental set of
particular solutions of the system (\ref{nonasys}) and
$k=(k_1,\ldots,k_n)$ is a set of $n$  arbitrary constants
associated with  each particular solution.
As a consequence of the
Implicit Function Theorem, the function $\Phi(x_{(1)},
\ldots,x_{(m)};\,\cdot\,):\mathbb{R}^n\to\mathbb{R}^n$ can be, at least locally
around generic points, inverted, so we can write
\begin{equation}
k=\Psi(x_{(0)}, \ldots,x_{(m)}) \label{defPsi}
\end{equation}
for a certain function $\Psi:{\mathbb{R}}^{n(m+1)}\to {\mathbb{R}}^n$.
Hereafter in order to handle a short notation we start writing $x_{(0)}$
instead of $x$. The  function $\Psi$, also called superposition function,
 provides us with a foliation which is invariant
under permutations of the $(m+1)$ variables. The functions  $\Phi$ and $\Psi$
are related by:
\begin{equation}
k=\Psi(\Phi(x_{(1)}, \ldots,x_{(m)};k_1,\ldots,k_n),x_{(1)}, \ldots,x_{(m)})\,.\label{implth}
\end{equation}
The fundamental property of the superposition function $\Psi$ is that as
\begin{equation}
k=\Psi(x_{(0)}(t),x_{(1)}(t), \ldots,x_{(m)}(t))\,,\label{kconstancy}
\end{equation}
the function  $\Psi(x_{(0)}, \ldots,x_{(m)})$ is constant on any
$(m+1)$-tuple of solutions of the system (\ref{nonasys}). This implies that the
 `diagonal prolongations' $\widetilde
Y(t,x_{(0)}, \ldots, x_{(m)})$ of the $t$-dependent  vector field
$Y(t,x)=Y^i(t,x)\,\partial/\partial x^i$, given by
$$\widetilde Y(t,x_{(0)}, \ldots,x_{(m)})=\sum_{a=0}^mY_a(t,x_{(a)})\,,\qquad
t\in {\mathbb{R}}\,,
$$
where
\begin{equation}\label{Ya}
Y_{a}(t,x_{(a)})=\sum_{i=1}^n Y^i(t,x_{(a)})\,\pd{}{x^i_{(a)}}
\end{equation}
are $t$-dependent  vector fields on ${\mathbb{R}}^{n(m+1)}$ 
which are tangent to the level sets of $\Psi$, i.e. the components $\Psi^i$ are
constants of motion. The level sets of $\Psi$
corresponding to
regular values define a $n$-codimensional foliation $\mathcal{F}$ on an open
dense subset $U\subset {\mathbb{R}}^{n(m+1)}$ 
and the family $\{\widetilde Y(t),\,t\in {\mathbb{R}}\}$ of vector
fields in ${\mathbb{R}}^{n(m+1)}$   consists of vector fields tangent to
the leaves of this foliation.

Remark that, as pointed out in \cite{CGM06}, for each  $(x_{(1)},\ldots,x_{(m)}) \in {\mathbb{R}}^{nm}$   there is one 
 point $(x_{(0)},x_{(1)},\ldots,x_{(m)})$ on the level
set $\mathcal{F}_k$ of this foliation $\mathcal{F}$ 
corresponding to $k=(k_1,\ldots,k_n)\in{\mathbb{R}}^n$,
namely,
$(\Phi(x_{(1)},\ldots,x_{(m)};k),x_{(1)},\ldots,x_{(m)})\in
\mathcal{F}_k $ (cf. (\ref{implth})); then, the projection onto the
last $m$ factors
$${\rm pr}:(x_{(0)},x_{(1)},\ldots,x_{(m)})\in {\mathbb{R}}^{n(m+1)}\mapsto
(x_{(1)},\ldots,x_{(m)})\in {\mathbb{R}}^{nm}
$$
induces diffeomorphisms on the leaves $\mathcal{F}_k$ of $\mathcal{F}$.
Such a foliation gives us
the superposition principle without referring to the function
$\Psi$: if we fix the point $x_{(0)}(0)$ 
(i.e. we choose a $k=(k_1,\ldots,k_n)$) and $m$ solutions
$x_{(1)}(t),\ldots,x_{(m)}(t)$, then $x_{(0)}(t)$ is the unique
point in ${\mathbb{R}}^n$ such that
$(x_{(0)}(t),x_{(1)}(t),\ldots,x_{(m)}(t))$ belongs to the same
leaf of $\mathcal{F}$ as
$(x_{(0)}(0),x_{(1)}(0),\ldots,x_{(m)}(0))$. This means that it is only
$\mathcal{F }$ really matters for the superposition rule.

Lie's main result \cite{LS} can be expressed as follows:

{\bf Theorem:}  {\it The system (\ref{nonasys}) on a differentiable manifold $N$
admits a superposition rule if and only if the $t$-dependent vector field
$Y(t,x)$ can be locally written in the form
$$Y(t,x)=\sum_{\alpha=1}^r b_\alpha(t)\, X_\alpha(x)$$
where the vector fields $X_\alpha$, $\alpha=1,\dots,
r$, close on a $r$-dimensional real Lie algebra, i.e. there
exist $r^3$ real numbers $c_{\alpha\beta}\,^\gamma $ such that
\begin{equation}
[X_\alpha,X_\beta]= \sum_{\gamma=1}^r c_{\alpha\beta}\,^\gamma\,
X_\gamma\,,\qquad \forall \alpha,\beta=1,\ldots,r\,.\label{Liesys}
\end{equation}
}

The number $m$ of solutions involved in the superposition rule for the
Lie system defined by (\ref{Liesys})
with generic $b_\alpha(t)$ is the
minimal $k$ such that the diagonal prolongations of
$X_1,\dots,X_r$ to $N^k$ are linearly independent at (generically)
each point: the only real numbers solution of the linear system
$$\sum_{\alpha=1}^r c_\alpha\, X_\alpha(x_{(a)})=0\,,\qquad a=1,\ldots ,k
$$
at a generic point $(x_{(1)},\dots,x_{(k)})$ is the trivial
solution $c_\alpha=0$, $\alpha=1,\ldots,m$,  for $k=m$, and there
are nontrivial solutions for $k<m$. Then, the superposition function $\Psi$ is 
made up of $t$-independent constants of motion for the prolonged $t$-dependent 
vector field $\widetilde Y$. We shall call them first integrals.

A possible  generalization consists on considering foliations which are not of codimension
$n$, or even choosing different Lie systems with the same associated Lie
algebra
for defining the prolonged vector field. These facts will be illustrated with
several examples.

\section{SODE Lie systems}

A system of second-order differential equations
$$\ddot x^i=f^i(t, x,\dot x)\,,\qquad i=1,\ldots ,n,$$ can be studied through
the
 system of first-order differential equations
$$\left\{
\begin{array}{rcl}
\dfrac{{dx^i}}{{dt}}&=&v^i\cr
\dfrac{{dv^i}}{{dt}}&=&f^i(t,x,v)
\end{array}
\right.
$$
with associated $t$-dependent vector field 
$$X=v^i\pd{}{x^i}+f^i(t,x,v)\pd{}{v^i}\,.$$

We call SODE Lie systems those for which $X$ is a Lie system, i.e. it can be
written as a linear combination with $t$-dependent coefficients of vector fields
closing a finite-dimensional real Lie algebra. There are many interesting
examples of such SODE Lie systems and next section is devoted to introduce some
particular examples.

\subsection{The 1-dim harmonic oscillator with time-dependent frequency}

 The equation of motion is $\ddot x= -\omega^2(t) x$,
with associated system
\begin{equation}
\left\{\begin{array}{rcl}\dot x&=&v\cr \dot v&=&-\omega^2(t)
    x\end{array}\right.\label{1dimho}
\end{equation}
and $t$-dependent vector field 
$$X=v\pd{}x -\omega^2(t) x\, \pd{}v\ ,
$$
which is a linear combination $X=X_2- \omega^2(t)X_1$, with 
$$X_1= x\pd{}v\,,\qquad X_2=v\, \pd{}x$$
such that 
$$[X_1,X_2]=2\, X_3\,, \quad [X_1,X_3]=- X_1 \,,\quad [X_2,X_3]=X_2\,,
$$
where $X_3$ is the vector field  given by 
$$X_3=\frac 12 \left(x\pd{}x-v\pd{}v\right) \,.
$$
Therefore $X$ defines a Lie system with associated  Lie algebra
$\goth{sl}(2,\mathbb{R})$. Actually, the vector fields $X_i$ are 
the fundamental vector fields corresponding to the
usual linear action and the basis of $\goth{sl}(2,\mathbb{R})$
given by the following traceless $2\times2$ real matrices
\begin{equation}
{\rm a}_1=\left(\begin{array}{cc}
0&0\\
-1&0
\end{array}\right),\quad
{\rm a}_2=\left(\begin{array}{cc}
0&-1\\
0&0
\end{array}\right),\quad
{\rm a}_3=\frac 12\left(\begin{array}{cc}
-1&0\\
0&1
\end{array}\right)\,.\label{basissl2}
\end{equation}

This system has no  first integrals, i.e. there are not $t$-independent
 constants of motion.

\subsection{The 2-dim isotropic harmonic oscillator with time-dependent frequency}

 The system of equations of motion is 
\begin{equation}
\left\{\begin{array}{rcl}\ddot x_1&=& -\omega^2(t) x_1\cr\ddot x_2&=& -\omega^2(t)
  x_2\end{array}\right.\label{2dimho}
\end{equation}
with associated system
$$\left\{
\begin{array}{rcl}\dot x_1&=&
v_1\cr \dot v_1&=&
-\omega^2(t) x_1\cr 
\dot x_2&=&
v_2\cr \dot v_2&=& -\omega^2(t) x_2
\end{array}\right.
$$
and $t$-dependent vector  vector field 
$$X=v_1\pd{}{x_1} -\omega^2(t) x_1\, \pd{}{v_1}+
v_2\pd{}{x_2} -\omega^2(t) x_2\, \pd{}{v_2}\ ,
$$
which is a linear combination, $X=X_2- \omega^2(t)X_1$, with 
$$X_1= x_1\pd{}{v_1}+ x_2\pd{}{v_2}
\,,\qquad X_2=v_1 \pd{}{x_1}+v_2 \pd{}{x_2}\,,
$$
such that
$$[X_1,X_2]=2\, X_3\,, \quad [X_1,X_3]=- X_1 \,,\quad [X_2,X_3]=X_2\,,
$$
where the vector field $X_3$ is defined by
$$X_3=\frac 12 \left(x_1\pd{}{x_1}-v_1\pd{}{v_1}+x_2\pd{}{x_2}
-v_2\,
\pd{}{v_2}
\right) \,.
$$

Once again $X$ defines a Lie system with associated  Lie algebra
 $\goth{sl}(2,\mathbb{R})$. However, in the preceding case there is no constant
 of motion and we shall see that in this last one $x_1v_2-x_2v_1$ is a constant
 of the motion.

\subsection{Milne--Pinney equation}

We call Milne--Pinney equation  the  second-order non-linear differential  equation
\cite{{P50},{Mil30}}:
\begin{equation}
\ddot x=-\omega^2(t)x+\frac k{x^3}\,,\label{Milneeq}
\end{equation}
where $k$ is a constant. It describes the time-evolution of an
 isotonic oscillator \cite{Cal69,{Pe90}}, i.e. an oscillator with inverse
 quadratic potential \cite{WS78}. This oscillator shares with the harmonic one
 the property of
having a period independent of the energy \cite{ChaVes05}, i.e. they are isochronous systems,
 and in the quantum case they have a equispaced spectrum \cite{ACMP07}.

The corresponding system of  first-order differential
equations is
$$\left\{
\begin{array}{rcl}
\dot x&=&v\cr
\dot v&=&-\omega^2(t)x+\dfrac{k}{{x^3}}\nonumber
\end{array}\right.
$$
and the associated $t$-dependent vector field
$$X=v\pd{}x+\left(-\omega^2(t)x+\frac k{x^3}\right)\pd{}v\,.
$$
This is a Lie system because it can be written as 
$$X=L_2-\omega^2(t)L_1\,,
$$
where
$$
L_1=x\frac {\partial}{\partial v},\qquad L_2=\frac k {x^3}\frac{\partial}{\partial v}
+v\frac{\partial}{\partial x}, 
$$
are such that 
$$
[L_1,L_2]=2L_3,\quad [L_3,L_2]=-L_2,\quad [L_3,L_1]=L_1
$$
with 
$$
 L_3=\frac 1 2 \left(x\frac{\partial}{\partial x}-v\frac{\partial}{\partial v}\right)\,,
$$
i.e. they  span a 3-dimensional real  Lie algebra $\goth{g}$
  isomorphic  to $\goth{sl}(2,\mathbb{R})$. Actually, one can show that they
  are the fundamental vector fields associated with the basis (\ref{basissl2})
relative to the following action of the group $SL(2,\mathbb{R})$   on a point
$(x_0,v_0)\in \mathbb{R}^2$:

If the matrix $A$ in  $SL(2,\mathbb{R})$ is given by 
$$A=\left(\begin{array}{cc}
\alpha\,&\,\beta\\ \gamma\,&\delta\,
\end{array}\right)\,,
$$
then
$\Phi(A,(x,v))=(\bar x,\bar v)$ with 

\begin{eqnarray}\label{ActPin}
\left\{\begin{aligned}
\bar x&={\rm sign}(x)\sqrt{\frac{k+\left[(\beta v+\alpha x)(\delta
      v+\gamma x)+ k({\delta\beta}/{x^2})\right]^2}{(\delta
    v+\gamma x)^2+k({\delta}/{x})^2}}\cr &\cr
\bar v&=\sqrt{\left(\delta v+\gamma x\right)^2+\frac{k\delta^2}{x^2}\left(1-\frac{x^2}{\delta^2\bar x^2}\right)}
\end{aligned}\right.
\end{eqnarray}
with sign$(x)={x}/{|x|}$.
\subsection{Ermakov system} 

Consider the system \cite{DL84,{PGL91}} 
\begin{equation}\label{Ermak}
\left\{\begin{array}{rcl}
\dot x&=&v_x\cr
\dot v_x&=&-\omega^2(t)x\cr
\dot y&=&v_y\cr
\dot v_y&=&-\omega^2(t)y+\dfrac {1}{{y^3}}\nonumber
\end{array}\right.
\end{equation}
with associated $t$-dependent vector field 
$$X=v_x\pd{}x+v_y\pd{}y-\omega^2(t)x\pd{}{v_x}+\left(-\omega^2(t)y+\frac 1
  {y^3}\right)\pd{}{v_y}\,,
$$
which is a linear combination with time-dependent coefficients,
 $X=-\omega^2(t)X_1+X_2$, of the vector fields 
$$X_1=x\pd{}{v_x}+y\pd{}{v_y}\,,\qquad X_2=v_x\pd{}x+v_y\pd{}y+\frac{1}
  {y^3}\pd{}{v_y}\,.
$$

This system is made up  by two Lie systems,  which correspond, respectively, to the
 examples of the 1-dimensional harmonic oscillator (\ref{1dimho}) and the Milne equation
 (\ref{Milneeq}),
 both  closing on
 a $\goth{sl}(2,\mathbb{R})$ algebra with $X_3$ given by 
$$X_3=\frac 12\left(x\frac{\partial}{\partial
    x}-v_x\frac{\partial}{\partial v_x}+y\frac{\partial}{\partial
    y}-v_y\frac{\partial}{\partial v_y}\right)\,.
$$

\subsection{Generalized Ermakov system}

It is the system given by \cite{{WS81},{RR79a},{RR79b},{GL94a},{R81},{RR80},{SaCa82}}:
\begin{eqnarray}\label{SOrder}
\left\{\begin{aligned}
\ddot{x}=\frac{1}{x^3}f(y/x)-\omega^2(t)x\cr
\ddot{y}=\frac{1}{y^3}g(y/x)-\omega^2(t)y
\end{aligned}\right.\nonumber
\end{eqnarray}
that for the choice  $f(u)=0$ and $g(u)=1$  reduces to the 
Ermakov system. 

This system of second-order equations can be written as one of first-order
equations
 by doubling the number of
degrees of freedom by introducing the new variables   $v_x$ and $v_y$:
\begin{equation}\label{FOrder}
\left\{\begin{aligned}
\dot x&=v_x\\
\dot v_x&=-\omega^2(t)x+\frac 1 {x^3} f( y/x)\\
\dot y&=v_y\\
\dot v_y&=-\omega^2(t)y+\frac 1 {y^3} g(y/x)\,.
\end{aligned}\right.\nonumber
\end{equation}
Such system  determines the integral curves of the vector field 
$$X=v_x\,\pd{}{x}+v_y\,\pd{}{v_y}+\left(-\omega^2(t)x+\frac 1 {x^3} f( y/
  x)\right)\pd{}{v_x}+\left(-\omega^2(t)y+\frac 1 {y^3} g(y/
    x)\right)\pd{}{v_y}\,,
$$
which can be written as a linear combination 
$$X=N_2-\omega^2(t)\, N_1
$$
where $N_1$ and $N_2$ are the vector fields 
$$
N_1=x\frac{\partial }{\partial v_x}+y\frac{\partial }{\partial v_y},\quad N_2=v_x\frac{\partial}{\partial x}+
\frac{1}{x^3}f(y/x)\frac{\partial}{\partial v_x}+v_y\frac{\partial}{\partial y}+
\frac{1}{y^3}g( y/x)\frac{\partial}{\partial v_y},
$$

Note that these vector fields generate a 3-dimensional real Lie algebra
with a third generator $$N_3=\frac 12\left(x\frac{\partial}{\partial
    x}-v_x\frac{\partial}{\partial v_x}+y\frac{\partial}{\partial
    y}-v_y\frac{\partial}{\partial v_y}\right)\,.$$ 
In fact, as  
$$
[N_1,N_2]=2N_3, \quad [N_3,N_1]=N_1, \quad  [N_3,N_2]=-N_2\,,
$$
they generate a Lie algebra isomorphic to   $\goth{sl}(2,\mathbb{R})$. Therefore the
system is a Lie system.    

The fact that the systems of preceding examples  are Lie systems with the same associated Lie algebra means that they can be
solved simultaneously in the group $SL(2,\mathbb{R})$ by the equation (see
e.g. \cite{CarRamGra,{CarRamcinc}})
$$\dot g\, g^{-1}=\omega^2(t)\, {\rm a}_1-{\rm a}_2\,,$$
where ${\rm a}_1$ and ${\rm a}_2$ are given in (\ref{basissl2}).

\section{The superposition functions for these examples}

Consider first the 1-dimensional harmonic oscillator (\ref{1dimho}). In order to look for a superposition rule we should
consider a system like in (\ref{2dimho}) and check whether the vector fields $X_1$
and $X_2$ are linearly independent in a generic point. There are many points in
which one can choose non trivial coefficients $\lambda_1,\lambda_2$ and $\lambda
_3$ such that $
\lambda_1\,X_1+\lambda_2\, X_2+\lambda_3\, X_3$ vanishes in such a  point.
On the contrary, if we introduce another copy and obtain the system (\ref{2dimho}),
only from the vanishing of such vector in a point such that 
$x_1\, v_2-x_2\, v_1=0$ and $x_3\, v_1-x_1\, v_2=0$ we cannot say that
$\lambda_1=\lambda_2=\lambda_3=0$, therefore $m=2$ and  consequently
 there is a superposition rule
involving two particular solutions.

Note that such system (\ref{2dimho}) admits a first integral  because the function  $F$ given by
$F(x_1,x_2,v_1,v_2)$ is such that
$X_2F=0$ iff there exists a function      $\bar F(\xi,v_1,v_2)$ with
$\xi=x_1v_2-x_2v_1$,
such that $F(x_1,x_2,v_1,v_2)=\bar F(\xi,v_1,v_2)$, and then from the second condition, 
$$x_1\, \pd{\bar F}{v_1}+x_2\, \pd{\bar F}{v_2}=0\,,
$$
we obtain the first integral, which corresponds to the angular momentum, 
$F(x_1,x_2,v_1,v_2)=x_1v_2-x_2v_1$,
which can be seen as a partial superposition rule. Actually, if $x_1(t)$ is a
solution of the first equation, then we obtain for each real number $k$
 the first-order differential
equation for the variable $x_2$ 
$$x_1(t)\, \frac{dx_2}{dt} =k+\dot x_1(t)x_2\,,
$$
from where $x_2$ can be found to be given by  
\begin{equation}
x_2(t)=k' x_1(t)+k\, x_1(t)\int^t\frac{d\zeta}{x_1^2(\zeta)}\,.\label{redunasol}
\end{equation}

In order to look for the superposition rule which does not involve quadratures 
 we should consider 
 three copies of the same oscillator,  and the extended vector fields $X_1$ and
$X_2$ given by 
$$X_1= x_1\pd{}{v_1}+ x_2\pd{}{v_2}+x  \pd{}{v  }\,,\qquad 
 X_2=v_1 \pd{}{x_1}+v_2 \pd{}{x_2}+v\pd {}x\ .
$$

We can determine the first integrals $F$ 
as solutions of $X_1F=X_2F=0$. The condition $X_2F=0$ says that there exists a
function $\bar F:{\mathbb{R}}^5\to {\mathbb{R}}^2$ such that $F(x_1,x_2,x,v_1,v_2,v)=\bar F (\xi_1
,\xi_2,v_1,v_2,v)$ with $\xi_1(x_1,x_2,x,v_1,v_2,v)=xv_1-x_1v$ and
$\xi_2(x_1,x_2,x,v_1,v_2,v)=xv_2-x_2v$, and the condition $X_1F=0$ transforms into 
$$x_1\pd{\bar F}{v_1}+x_2\pd{\bar F}{v_2}+x\pd{\bar F}{v}=0\,,
$$ 
i.e. $\xi_1$ and $\xi_2$ are first integrals (Of course, $\xi=x_1v_2-x_2v_1$ is
also a first integral.They produce a superposition
rule, because from
$$\left\{\begin{array}{crl} xv_2-x_2v&=&k_1\cr x_1v-v_1x&=&k_2\end{array}\right.
$$ 
 we obtain the expected superposition rule for two solutions:
$$x=c_1\, x_1+c_2\, x_2\,,\qquad v=c_1\, v_1+c_2\, v_2\,,\qquad
c_i=\frac{k_1}{k}, \  k=x_1v_2-x_2v_1\,.
$$

As a second example, consider the Milne system given in (\ref{Milneeq})
The generators of this Lie system
with algebra   $\goth{sl}(2,\mathbb{R})$  span a distribution of dimension two and there
is  no first integral of the  motion for such subsystem. By adding the 
other $\goth{sl}(2,\mathbb{R})$ linear Lie system appearing in the Ermakov
system, 
the harmonic oscillator with time
dependent angular frequency, as  the distribution in the 4-dimensional 
space  is of rank three, there is an integral of motion.  The first integral can
be obtained 
from $L_1F=L_2F=0$. But $L_1F$ means that $F(x,y,v_x,v_y)=\bar F(x,y,\xi)$ with $\xi=xv_y-yv_x$,
and then $L_2F=0$ is written
$$
v_x\pd{\bar F}{x}+v_x\pd{\bar F}{x}+\frac x{y^3}\pd{\bar F}{\xi}
$$ and we obtain the associated system of characteristics  
 $$\frac {x\,dy-y\, dx}{\xi}=\frac {y^3\,d\xi}{x}\Longrightarrow
  \frac{d(x/y)}{\xi}+\frac{y\,d\xi}{x}=0\,,
$$
from where the following first integral is found \cite{Le67}:
$$
\psi(x,y,v_x,v_y)=\left(\frac{x}{y}\right)^2+\xi^2=\left(\frac{x}{y}\right)^2+
(xv_y-yv_x)^2\,,
$$
which is the well-known Lewis--Ermakov invariant \cite{DL84,{PGL91},{RR79a}}. 

We can follow a similar path in the case of  the generalized Ermakov system.
There exists  a first  integral for the motion, $F:\mathbb{R}^4\rightarrow \mathbb{R}$,
for any $\omega^2(t)$, because 
 this Lie system has an associated integrable distribution of rank three
 and the manifold is
 4-dimensional. 

This first integral $F$  satisfies 
$N_iF=0$ for $i=1,\ldots,3$, but as $[N_1,N_2]=2N_3$ it is enough to impose
$N_1F=N_2F=0$. Then, if $N_1F=0$, 
$$
x\,\pd{F}{v_x}+y\, \pd{F}{v_y}=0\,,
$$
and according to the method of
characteristics we obtain:
$$
\frac{dx}{0}=\frac{dy}{0}=\frac{dv_x}{x}=\frac{dv_y}{y}
$$
and therefore there exists a function $\bar F:\mathbb{R}^3\rightarrow \mathbb{R}$ such that
$F(x,y,v_x,v_y) =\bar F(x,y,\xi=xv_y-yv_x)$.  The
condition   $N_2F=0$ reads now 
$$
v_x\frac{\partial\bar F}{\partial x}+v_y\frac{\partial\bar F}{\partial
  y}+\left(-\frac{y}{x^3}f({y}/{x})+
\frac{x}{y^3}g({y}/{x})\right)\frac{\partial \bar F}{\partial \xi}\,.
$$

We can therefore consider the associated system of 
 the characteristics:
$$
\frac{dx}{v_x}=\frac{dy}{v_y}=\frac{d\xi}{-\frac{y}{x^3}f({y}/{x})+\frac{x}{y^3}g({y}/{x})}
$$
But using that 
$$
\frac{-y\,dx+x\,dy}{\xi}=\frac{dx}{v_x}=\frac{dy}{v_y}\,,
$$
we arrive to 
$$
\frac{-y\,dx+x\,dy}{\xi}=\frac{d\xi}{-\frac{y}{x^3}f(\frac{y}{x})+\frac{x}{y^3}g(\frac{y}{x})}
$$
i.e.
$$
-\frac{y^2d\left(\frac{x}{y}\right)}{\xi}=
\frac{d\xi}{-\frac{y}{x^3}f(\frac{y}{x})+\frac{x}{y^3}g(\frac{y}{x})}\\
$$
and integrating we obtain the following first-integral, with $u=y/x$, 
$$
\frac 12 \xi^2+\int^u\left[-\frac 1{\zeta^3}\, f\left(\frac 1\zeta\right)+
  \zeta\,g\left(\frac 1\zeta
\right)\right]\,d\zeta=C\,.
$$

 This
first  integral allows us  to determine, by means of quadratures,  a solution of one subsystem 
 in terms of  a solution of the other equation.

\subsection{\bf The Pinney equation revisited}

We mentioned before the possibility of obtaining solutions of a given system
from particular solutions of another related system. We next study a
 particular example, studied by Pinney long time ago \cite{P50}.
 Consider the system of first-order differential equations:
$$
\left\{
\begin{array}{rcl}
\dot x&=&v_x\cr
\dot y&=&v_y\cr
\dot z&=&v_z\cr
\dot v_x&=&-\omega^2(t)x+\dfrac{k}{{x^3}}\cr
\dot v_y&=&-\omega^2(t)y\cr
\dot v_z&=&-\omega^2(t)z
\end{array}\right.
$$
which corresponds to the vector field 
$$X=v_x\pd{}x+v_y\pd{}y+v_z\pd{}z+\frac{k}{x^3}
\frac{\partial}{\partial v_x}-\omega^2(t)\left(x\frac{\partial }{\partial v_x}+y\frac{\partial }{\partial v_y}+
z\pd{}{v_z}\right)\,.
$$
The vector field $X$ can be expressed as $X=N_2-\omega^2(t)N_1$ where the vector
fields $N_1$ and $N_2$ are:
$$
N_1=y\frac{\partial }{\partial v_y}+x\frac{\partial }{\partial v_x}+
z\pd{}{v_z},\quad N_2=v_y\frac{\partial}{\partial y}+
\frac{1}{x^3}\frac{\partial}{\partial v_x}+v_x\frac{\partial}{\partial
  x}+v_z\frac{\partial}{\partial z},
$$ 
These vector fields generate a 3-dimensional real Lie algebra with the
vector field $N_3$ given by 
$$
 N_3=\frac 12\left(x\frac{\partial}{\partial x}-v_x\frac{\partial}{\partial
     v_x}+y\frac{\partial}{\partial y}-v_y\frac{\partial}{\partial
     v_y}+z\frac{\partial}{\partial z}-v_z\frac{\partial}{\partial
     v_z}\right)\,.
$$
In fact, they generate a Lie algebra isomorphic to   $\goth{sl}(2,\mathbb{R})$ because  
$$
[N_1,N_2]=2N_3, \quad [N_3,N_1]=N_1, \quad  [N_3,N_2]=-N_2\,.
$$

The distribution generated by these fundamental vector fields has rank three. 
Thus, as the manifold of the Lie system is of dimension six we obtain three
 time-independent integrals of motion. 

\begin{itemize}

\item The Ermakov invariant
 $I_1$ of the subsystem involving variables $x$ and $y$.

\item The Ermakov invariant $I_2$ 
of the subsystem involving variables $x$ and $z$ 

\item  The Wronskian $W$ of the subsystem involving variables $y$ and $z$
 has

\end{itemize}

They define a  foliation  with 3-dimensional leaves. 
We can use this foliation for obtaining in terms of them such  a superposition rule.

The Ermakov invariants read as:
$$
I_1=\frac 12\left((yv_{x}-xv_y)^2+k\left(\frac {y}x\right)^2\right)\,,\qquad
I_2=\frac 12\left((xv_{z}-zv_x)^2+k\left(\frac {z}x\right)^2\right)\,,
$$
and $W$ is:
$$
W=yv_{z}-zv_{y}\,.
$$
In terms of these three integrals we can obtain an explicit expression of $x$ in terms of $y, z$ and the integrals $I_1, I_2, W$:
$$
x=\frac {\sqrt 2} W\left(I_2y^2+I_1z^2\pm\sqrt{4I_1I_2-kW^2}\ yz\right)^{1/2}\,.
$$	
This can be interpreted, as pointed out by Pinney \cite{P50}, 
 as saying that there is a superposition rule allowing
us to express  the general solution of the Milne--Pinney equation in terms of two
independent solutions of the corresponding harmonic oscillator with
the same  time-dependent angular frequency.

\section{The reduction technique for SODE Lie systems}

We aim to illustrate by simple examples the usefulness of the reduction
technique for Lie systems in this particular case of SODE systems. The main
point is that given a Lie system in a homogeneous space for the associated
group $G$, the knowledge of a particular solution allows us to reduce the
problem to a new Lie system in the corresponding stability  subgroup
\cite{CarRamGra,{CarRamcinc}}.
Actually, if the Lie system is defined in the Lie group $G$ by 
the equation 
\begin{equation}
\dot g(t)\,g^{-1}(t)={\rm a}(t)\,,\qquad g(0)=e\label{eqinG}
\end{equation} and we know a particular solution 
$x_1(t)$ of 
the associated system in a homogeneous space, we can choose a curve $\bar g(t)$  in $G$ such
that $\bar g(0)=e$ and $\Phi(\bar g(t),x_0)=x(t)$, and therefore, as also $\Phi(g(t),x_0)=x(t)$,
there is a curve $h(t)$ in $G_{x_o}$ such that $g(t)=\bar g(t)h(t)$. Then
$h(t)$ is the solution of the Lie system in $G_{x_0}$
$$\dot h(t) \, h^{-1}(t)={\rm Ad\,}({\bar{g}}^{-1}(t))({\rm a}(t))+\dot {\bar g}^{-1}(t) \, {\bar g}(t)\,,\qquad h(0)=e\,.
$$
Once such a Lie system has been solved the solution in the group is $g(t)=\bar
g(t)h(t)$ and therefore the solution starting from any point $y_0$ of the associated
Lie system in a homogeneous
space is given by $\Phi(\bar g(t)h(t),y_0)$. As it is shown in the reduction technique, the transformation through the
curve $\bar g(t)$ reduces the problem from the group $G$ to the stability group of the initial condition of the particular solution
$x_0$. 

Instead, we can be interested in the reduction of the Lie system to one in the stability
subgroup of a previously fixed point $z_0$ and we should proceed as follows. Let
$g_0$ be a fixed element in $G$ such that $\Phi(g_0,x_0)=z_0$. Then $g(t)$ is a
solution of (\ref{eqinG}) iff $\widetilde g(t)=g(t)\,g_0^{-1} $ is a solution of the
same equation as in (\ref{eqinG}) but with $\widetilde g(0)=g_0^{-1}$. Moreover, such
curve $\widetilde g(t)$ can also be used to define the solution 
of the system in the
homogeneous space starting from $x_0$ by means of $\Phi(\widetilde g(t),z_0)$. 
Note that if $x_1(t) $ is a
particular solution in the given homogeneous space with initial condition
$x_1(0)=x_0$,
 we can choose a curve
$g_1(t)$ in $G$ such that 
$\Phi(g_1(t),z_0)=x_1(t)$ and then from  $\Phi(g_1(t),z_0)=\Phi(\tilde
g(t),z_0)=x_1(t)$  we obtain that:
$$
z_0=\Phi(g_1^{-1}(t),\Phi(g_1(t),z_0))=\Phi(g_1^{-1}(t),\Phi(\tilde g(t),z_0))=\Phi(g_1^{-1}(t)\tilde g(t),z_0).
$$
Thus, $h(t)=g^{-1}_1(t)\tilde g(t)$ lies in  the stability group of $z_0$,
$G_{z_0}$,
 and, consequently, $g_1(t)$ and  $\tilde g(t)$ differ on a curve in
 $G_{z_0}$. Recall that $\tilde g(t)$ satisfies the equation (\ref{eqinG}) but
 with $\tilde g(0)=g_0^{-1}$ and therefore $h(t)$ is such that:
\begin{eqnarray}\label{tr}
\dot h(t)h^{-1}&=&\frac{d}{dt}(g_1^{-1}(t)\tilde g(t))(g_1^{-1}(t)\tilde g(t))^{-1}\cr&=&\dot g^{-1}_1(t)g_1(t)+{\rm Ad}(g^{-1}_1(t))({\rm a}(t))={\rm a}'(t)\in
T_IG_{z_0},
\end{eqnarray}
and satisfies the initial condition $h(0)=g_1^{-1}(0)g_0^{-1}$.

Then, the transformation given by $g_1^{-1}(t)$ changes the initial equation in
$G$ associated to ${\rm a}(t)$ into a new equation in $G_{z_0}$ associated with
${\rm a}'(t)$ independently of the initial condition of the particular solution
$x_1(t)$. When such stability group  $G_{z_0}$ has a solvable Lie algebra
$\goth{g}_{z_0}$ we can solve the equation in $G_{z_0}$ for any ${\rm a}'(t)$.

Conversely, if $h(t)$ is a curve in $G_{z_0}$ solution of the Lie system in $G_{z_0}$ 
\begin{equation}
\dot h(t)h^{-1}(t)={\rm a}'(t),\label{eqinh}
\end{equation}
satisfying the initial condition $h(0)=g_1^{-1}(0)g_0^{-1}$, then
$g(t)=g_1(t)\, h(t)\, g_0$ is the curve solution of our initial Lie system.
On the other side, the equations defining Lie systems are right-invariant and
therefore if $h'(t)$ is a curve solution of (\ref{eqinh}) but with the usual
initial condition $h'(0)=e$, then the solution of (\ref{eqinh}) with initial
condition $h(0)=g_1^{-1}(0)g_0^{-1}$ is $h(t)=h'(t)g_1^{-1}(0)g_0^{-1}$.
Therefore, the solution of the initial Lie system is given by 
$g(t)=g_1(t)h'(t)g_1^{-1}(0)$.

Let us apply this theoretical development to three particular cases of this
reduction
procedure: a reduction of a harmonic  oscillator with a time-dependent frequency using
one particular solution, a reduction of a Milney--Pinney equation through a
particular solution, and a reduction of a Milney--Pinney equation by means
of  a particular solution of a harmonic oscillator with  the same 
time-dependent frequency.

Consider first the simple example  (\ref{1dimho}) of the harmonic oscillator with a
time-dependent frequency and assume that a particular solution $x_1(t)$ is known.
This  is a Lie system in $\mathbb{R}^2$ which is a homogeneous
space for the group $SL(2,\mathbb{R})$. The orbit of the point $(1,0)$ 
under the
linear action is all $\mathbb{R}^2$,  the stability group of such point being the 
1-dimensional Lie subgroup generated by $v\, \partial/\partial x$. We can
choose the matrix
\begin{equation} 
g_1(t)=\matriz{cc}{x_1(t)&0\cr\dot x_1(t)&x_1^{-1}(t)}\label{curvag1}
\end{equation}
as a curve mapping the point $(1,0)$ onto the given solution. Then we should
make the change of coordinates corresponding to the transformation in the homogeneous manifold induced by $g^{-1}_1(t)$:
$$\matriz{c}{x\cr v_x}=\matriz{cc}{x_1(t)&0\cr\dot
  x_1(t)&x_1^{-1}(t)}\matriz{c}{z\cr v_z}\,,
$$
and then we obtain the new second-order differential equation which is a
first-order 
system in $\dot z$,
$$x_1(t)\, \ddot z+2\, \dot x_1(t)\, \dot z=0\,,
$$ and allows us to find the general solution by means of
a quadrature, because if $u=\dot z$, the general solution of 
 $x_1(t)\, \dot u+2\, \dot x_1(t)\, u=0$ is $u=k/(x_1(t))^2$, for any constant
and then we obtain by a second quadrature the expression (\ref{redunasol}). The fact
that the new equation is solvable is related to the fact that the new equation is related with
the solvability of $\goth{g}_{z_0}$. This allows us to obtain its solutions by quadratures.
Also, note that this change $x=x_1(t)\, z$ corresponds to the traditional
d'Alembert method of reduction of order (see e.g. \cite{Rain}). 

Consider now the example (\ref{Milneeq}) of the Milne--Pinney equation. This is a homogeneous
case if 
we consider that either  $x>0$ or $x<0$ and  we restrict ourselves to one
 of these cases. Now, given a particular solution $x_1(t)$ of the Milne--Pinney
 equation the $g_1(t)$ constructed with $x_1(t)$ as in (\ref{curvag1})
 transforms 
the point $z_0=(1,0)$ into a solution $(x_1(t),\dot x_1(t))$ of this
differential equation. Thus, as in the last case,  we can use
again $g_1^{-1}(t)$  to transform the initial equation in the group 
given by ${\rm a}(t)$ into a new one given by ${\rm a}'(t)$ that actually is a
Lie equation  in $\goth{g}_{z_0}$. This new one will be solvable because $\goth{g}_{z_0}$ is solvable in this case. 

In this way, we start with:
\begin{equation}
\Phi(g_1^{-1}(t),(x,v_x))=(z,v_z).
\end{equation}
The new equation in the group determined by the curve ${\rm a}'(t)$ constructed
with $g_1^{-1}(t)$,  
\begin{equation}
\dot h(t) \, h^{-1}(t)={\rm Ad\,}(g_1^{-1}(t))({\rm a}(t))+ \dot{g}_1^{-1} \, g_1(t)={\rm a}'(t)\,,
\end{equation}
turns out to be in this case: 
$$\dot h(t) \, h^{-1}(t)={\rm a}'(t)=\frac 1{x_1^2(t)}(k\,{\rm a_1}-{\rm
  a}_2)\,.$$ 
If we consider a new variable $\tau$ defined by 
\begin{equation}
 \tau=\int_0^t\frac {d\zeta}{x^2_1(\zeta)}\,,\label{timetau}
\end{equation}
the new equation in the group is given by:
\begin{equation}
\frac{dh(\tau)}{d\tau}h^{-1}(\tau)=k\,{\rm a_1}-{\rm a}_2\,,
\end{equation}
then the solution with $h(0)=e$ is $h(\tau)=\exp((k{\rm a}_1-{\rm a}_2)\tau)$
and we arrive to the solution for the original Lie system
$g(t)=g_1(t)h(\tau)g_1^{-1}(0)$. Thus, the solution for the 
time evolution is:
\begin{equation}
\begin{aligned}
g(t)&=
\left(\begin{array}{cc}
x_1(t)& 0\\
\dot x_1(t)&x_1^{-1}(t)
\end{array}\right)
\left(\begin{array}{cc}
\cos (\sqrt{k}\tau)& \frac{1}{\sqrt{k}}\sin(\sqrt{k}\tau)\\
-{\sqrt{k}}\sin(\sqrt{k}\tau) & \cos(\sqrt{k}\tau)
\end{array}\right)
\left(\begin{array}{cc}
x_1^{-1}(0) &0\\
-\dot x_1(0)&x_1(0)
\end{array}
\right)\\
&=
\left(
\begin{array}{cc}
 x_1(t)\cos \left(\sqrt{k} \tau \right) & \frac{x_1(t) \sin \left(\sqrt{k} \tau \right)}{\sqrt{k}} \\
- \frac{\sqrt{k} \sin \left(\sqrt{k} \tau \right)}{x_1(t)}+\cos \left(\sqrt{k} \tau \right) \dot x_1(t) & \frac{\cos\left(\sqrt{k} \tau \right)}{x_1(t)}+\frac{\sin\left(\sqrt{k}\tau \right) \dot x_1(t)}{\sqrt{k}}
\end{array}
\right)
\left(\begin{array}{cc}
x_1^{-1}(0) &0\\
-\dot x_1(0)&x_1(0)
\end{array}
\right).\nonumber \\
\end{aligned}
\end{equation}
Now, if $\Phi$ denotes the action for the Milne--Pinney equation and we define:
\begin{equation}
\left(\begin{array}{c}
A\\B
\end{array}\right)=\Phi\left(
\left(\begin{array}{cc}
x_1^{-1}(0) &0\\
-\dot x_1(0)&x_1(0)
\end{array}
\right),\left(
\begin{array}{c}
x_0\\
v_0\end{array}\right)\right)
\end{equation}
then the set of solutions of the Milne--Pinney equation is written  in terms of $A, B$  as:
{\small \begin{equation*}
x(t)=\sqrt{\frac{(B^2+\frac{k}{A^2}+A^2k+(-B^2-\frac{k}{A^2}+A^2)\cos(2\sqrt{k}\tau(t))+2AB\sqrt{k}\sin(2
\sqrt{k}\tau(t))x_1^2(t)}{2k}}
\end{equation*}}

As a final example,  consider  once again the Milne--Pinney equation but assume that
a particular solution $x_1(t)$ of the time-dependent frequency harmonic
oscillator   with the same frequency as the Milne--Pinney equation is known.
 Recall that it has been shown that both equations are related with the same Lie
 system
 in the same group $G$. In this case, once again $g_1(t)$ given by
 (\ref{curvag1}) transforms $(1,0)$ into a particular solution $(x_1(t),\dot
 x_1(t))$ of
 the time-dependent harmonic oscillator as a first-order differential
 equation. 
 Thus, the transformation induced in $G$ changes the initial differential
 equation in
 $G$ corresponding to the harmonic and the Pinney equation, which is  the same
 one,
 into a new one in the stability group $G_{z_0}$. Now, the new equation in the group $G_{z_0}$ is given by:
\begin{equation}
\dot h(t)h^{-1}(t)={\rm Ad\,}(g_1^{-1}(t))({\rm a}(t))+\dot{g}_1(t) \, g_1^{-1}(t)=-\frac{1}{x_1^2(t)}{\rm a}_2
\end{equation}
and making use of a 
reparametrization given by (\ref{timetau}) we obtain that the new equation is:
\begin{equation}
\frac{dh(\tau)}{d\tau}\,h^{-1}(\tau)=-{\rm a}_2,
\end{equation}
and the solution with $h(0)=e$ is:
\begin{equation}
h(\tau)=\left(\begin{array}{cc}
1\,&\,\tau\\
0\,&\,1
\end{array}\right),
\end{equation}
from where we obtain the solution of our Lie system:
\begin{equation}
\begin{aligned}
g(t)&=
\left(\begin{array}{cc}
x_1(t)& 0\\
\dot x_1(t)&x_1^{-1}(t)
\end{array}\right)
\left(\begin{array}{cc}
1& \tau\\
0 & 1
\end{array}\right)
\left(\begin{array}{cc}
x_1^{-1}(0) &0\\
-\dot x_1(0)&x_1(0)
\end{array}
\right)\\
&=
\left(\begin{array}{cc}
x_1(t) & x_1(t)\tau\\
\dot x_1(t) &\dot x_1(t)\tau +x^{-1}_1(t)
\end{array}\right)
\left(\begin{array}{cc}
1/x_1(0) &0\\
-\dot x_1(0)&x_1(0)
\end{array}
\right)\\
&=\left(\begin{array}{cc}
x_1(t)/x_1(0) -x_1(t)\dot x_1(0)\tau&x_1(0)x_1(t)\tau\\
\dot x_1(t)/x_1(0)-\dot x_1(0)/x_1(t)-\tau \dot x_1(t) \dot x_1(0)&x_1(0)(\dot x_1(t)\tau + x_1^{-1}(t))
\end{array}
\right).\nonumber\\
\end{aligned}
\end{equation}
If we introduce the parameters $A$ and $B$ by
\begin{equation}
\left(\begin{array}{c}
A\\B
\end{array}\right)=\Phi\left(
\left(\begin{array}{cc}
x_1^{-1}(0) &0\\
-\dot x_1(0)&x_1(0)
\end{array}
\right),\left(
\begin{array}{c}
x_0\\
v_0\end{array}\right)\right).
\end{equation}
where $\Phi$ denotes  the action for the Milne--Pinney equation, we obtain:

\begin{equation*}
x(t)=\frac{x_1(t)}{A}\sqrt{A^4+2 A^3 B \tau(t)+ (A^2 B^2+k)\tau(t)^2}
\end{equation*}

\section*{Acknowledgements}
 Partial financial support by research projects MTM2006-10531 and E24/1 (DGA)
and a F.P.U. grant from  Ministerio de Educaci\'on y Ciencia are acknowledged.


\begin{thebibliography}{AAAAA}

\bibitem
{LS}  S. Lie,
{\sl Vorlesungen \"uber continuierliche Gruppen mit 
Geometrischen und anderen Anwendungen\/},
Edited and revised by G. Scheffers,
Teubner, Leipzig, 1893.
\bibitem{NI1} N.H. Ibragimov,  {\sl Primer of group analysis}, Znanie, No. 8, Moscow,
1989. (Russian). Revised edition in English: Introduction to modern
group analysis, Tau, Ufa, 2000.
\bibitem{NI2} N.H. Ibragimov, {\sl Elementary Lie group analysis and ordinary
    differential equations}, J. Wiley, Chichester, 1999.
\bibitem{Win83}
{P. Winternitz},
{\it Lie groups and solutions of nonlinear differential equations},
 in: {\sl Nonlinear Phenomena}, K.B. Wolf Ed., Lecture Notes in Physics {\bf 189},
{Springer-Verlag, N.Y., 1983}
\bibitem{CGM}  
J.F. Cari\~nena,  J. Grabowski and G. Marmo,
{\sl Lie--Scheffers systems: a geometric approach\/},
Bibliopolis, Napoli, 2000.
\bibitem{CarRamGra}  
J.F. Cari\~nena, J. Grabowski and  A. Ramos,
{\it Reduction of time-dependent systems admitting a
 superposition principle}, 
 Acta Appl.  Math. {\bf 66}, 67--87 (2001).
\bibitem{CGM01} J.F. Cari\~nena,  J. Grabowski and G. Marmo,
{\it Some applications in physics of differential equation systems admitting a superposition rule},
 Rep. Math. Phys. {\bf 48},   47--58   (2001).
\bibitem{And80}
R.L. Anderson, {\it  A nonlinear superposition principle admitted by coupled Riccati equations of the projective type},  
 Lett. Math. Phys. {\bf 4}, 1--7 (1980).
\bibitem{HarWinAnd83}
J. Harnad, P. Winternitz and R.L. Anderson,  
{\it Superposition principles for matrix Riccati equations},
J. Math. Phys. {\bf 24}, 1062--72 (1983).
\bibitem{OlmRodWin8687}
M.A. del Olmo, M.A. Rodr\'{\i}guez and P. Winternitz,
{\it Simple subgroups of simple Lie groups and 
nonlinear differential equations with superposition principles},
 J. Math. Phys. {\bf 27}, 14--23 (1986); {\bf 28},  530--5 (1987).  
\bibitem{CarRamdos}  
J.F. Cari\~nena  and A. Ramos,
{\it Riccati equation, Factorization Method and Shape Invariance\/},
Rev. Math. Phys.  {\bf A 12}, 1279--304 (2000).
\bibitem{CarRamcinc} 
J.F. Cari\~nena and  A. Ramos,
{\it A new geometric approach to Lie systems and physical applications},
{Acta Appl.  Math.} {\bf 70},   43--69  (2002).
\bibitem{CarMarNas}  
J.F. Cari\~nena, G. Marmo and J. Nasarre,
{\it  The nonlinear superposition principle and the Wei--Norman method},
 Int. J. Mod. Phys.  {\bf A 13}, 3601--27 (1998).
\bibitem{CGM06}  
J.F. Cari\~nena,  J. Grabowski and G. Marmo,
{\it Superposition rules, Lie theorem and partial differential equations},
math-ph/0610013
\bibitem{ermakov} V. Ermakov, {\it Second order differential
  equations. Conditions of complete integrability}, Univ. Isz. Kiev Series III
{\bf 9},  1--25  (1880) (translation by A.O. Harin). 
\bibitem{P50} E. Pinney, {\it The nonlinear differential equation}
$y''+p(x_1)y'+cy^{-3}=0$, Proc. A.M.S. {\bf 1}, 681 (1950).
\bibitem{RR80b} J.L. Reid   and J.R. Ray, {\it Ermakov systems, nonlinear
superposition and solutions of nonlinear equations of motion},
J. Math. Phys. {\bf 21}, 1583--87 (1980).
\bibitem{WS81} W. Sarlet, {\it Further generalization of Ray--Reid systems},
  Phys. Lett. {\bf A 82} 161--64 (1981).
\bibitem{Le67} H.R. Lewis, {\it Classical and Quantum Systems with Time-Dependent
  Harmonic-Oscillator-Type Hamiltonians}, Phys. Rev. Lett. {\bf 18}, 510 - 512 (1967)
\bibitem{Mil30}  W.E. Milne, {\it The numerical determination of characteristic
  numbers}, Phys. Rev. {\bf 35}, 863--67 (1930)
\bibitem{Cal69} {F. Calogero},
{\it  Solution of a three body problem in one dimension}, {\sl J.
Math. Phys.} {\bf  10},  2191--2196  (1969).
\bibitem{Pe90}{A.M. Perelomov},
{\sl ``Integrable systems of classical mechanics and Lie algebras"},
(Birk\-hauser, 1990).
\bibitem{WS78} W. Sarlet, {\it Exact invariants for time-dependent Hamiltonian
    systems with one degree of freedom}, J. Phys. A:Math. Gen. {\bf 11},
  843--54 (1978)
\bibitem{ChaVes05} {O.A. Chalykh and A.P. Vesselov}, {\it  A remark
on rational isochronous potentials}, {J. Nonlin. Math.
Phys.} {\bf  12} (Suppl. 1),    179--183 (2005).
\bibitem{ACMP07} {M. Asorey, J.F. Cari\~nena , G. Marmo and A.M.  Perelomov},
{\it  Isoperiodic classical systems and their quantum
counterparts}, {Ann. Phys.} {\bf 322}, (to appear, 2007).
\bibitem{DL84} A.K. Dhara  and S.V. Lawande, {\it Time-dependent invariant sand the
  Feynman propagator}, Phys. Rev. {\bf A 30}, 560-7 (1984).
\bibitem{PGL91}P.G.L. Leach, {\it Generalized Ermakov systems}, 
 Phys. Lett. {\bf 158 A}, 102--06 (1991).
\bibitem{RR79a} J.R. Ray  and J.L. Reid,
{\it  More exact invariants for the
  time-dependent harmonic oscillator}, Phys. Lett. {\bf 71 A}, 317--18 (1979)
\bibitem{RR79b}J.R. Ray  and J.L. Reid, {\it Exact time-dependent invariants for
$N$-dimensional systems}, Phys. Lett. {\bf 74 A}, 23--25 (1979)

\bibitem{GL94a}K.S. Govinder  and  P.G.L. Leach, {\it Ermakov systems: a group
  theoretic approach},  Phys. Lett. {\bf 186 A}, 391--95 (1994).
\bibitem{R81}  J.R. Ray, {\it Invariants for nonlinear equations of motion}, 
Progr. Theor. Phys. {\bf 65}, 877--82 (1981).
\bibitem{RR80}
J.L. Reid and J. R. Ray,
{\it Ermakov systems, Noether's theorem and the Sarlet-Bahar method},
Lett. Math. Phys. {\bf 4}, 235--240 (1980).
\bibitem{SaCa82}
W. Sarlet and F. Cantrijn,
{\it A generalization of the nonlinear superposition idea for Ermakov systems},
Phys. Lett. {\bf 88 A}, 383--387 (1982).
\bibitem{Rain} E.D. Rainville, {\sl Elementary Differential Equations},
  Macmillan, 1974.
\end{thebibliography}
\end{document}